\documentclass[apl,reprint]{revtex4-1}

\usepackage{graphicx}
\usepackage{amssymb}
\usepackage{amsmath}
\usepackage{natbib}
\usepackage{color}   

\newcommand{\I}{\mathrm{i}}

\begin{document}

\title{Ultra-thin low frequency perfect sound absorber with high ratio of active area} 

\author{Yves Aur\'egan}\email{yves.auregan@univ-lemans.fr}
\affiliation{Laboratoire d'Acoustique de l'Universit\'e du Mans, Centre National de la Recherche Scientifique (CNRS), Le Mans Universit\'e, Avenue Olivier Messiaen, 72085 Le Mans Cedex 9, France}

\begin{abstract}

A concept of ultra-thin low frequency perfect sound absorber is proposed and demonstrated experimentally. 
To minimize non-linear effects, an high ratio of active area to total area is used to avoid large localized amplitudes. 
The absorber consists of three elements: a mass supported by a very flexible membrane, a cavity and a resistive layer.  
The resonance frequency of the sound absorber can be easily adjusted just by changing the mass or thickness of the cavity.  
A very large ratio between wavelength and material thickness is measured for a manufactured perfect absorber (ratio = 201) . 
It is shown that this high sub-wavelength ratio is associated with narrowband effects and that the increase in the sub-wavelength ratio is limited by the damping in the system.

\end{abstract}

\maketitle

Sound absorption is of great interest to engineers who want to reduce sound reflection in room acoustics or reduce sound emissions. There is a need for innovative acoustic absorbent materials, effective in low frequencies while being able to deal with spatial constraints present in real applications.
Innovative ultra-thin materials are also useful tools for the scientific community to manipulate sound waves and obtain negative refraction \cite{cummer2016controlling,kaina2015negative}, sub-wavelength imaging \cite{zhu2011holey,qi2018ultrathin}, cloaking \cite{faure2016experiments}, etc. 

Traditional sound absorption structures use perforated and micro-perforated panels covering air or porous materials.  \cite{allard2009,maa1998}. 
These materials have a low reflection of normal incident waves at frequencies such that the wavelength ($\lambda = c_0/f$ where $c_0$ and $f$ are the sound velocity in air and the frequency) is about four times the thickness of the material $H$ leading to a sub-wavelength ratio $r_H = \lambda/H \simeq 4$.  
There has been a very significant reduction in the thickness of the absorbent materials \cite{yang2017optimal} by using space-coiling structures \cite{cai2014,chen2017,wang2018,long2018multiband} of by using slow-sound inside the material \cite{groby2015,yang2016,jimenez2017}. 

Nevertheless, all these structures have a front plate with small holes leading to a very low open area ratio ($\sigma$ = area of the orifices on the total area). 
In this case, since the velocity in the orifices is equal to the acoustic velocity in the incident wave divided by $\sigma$, some non-linear effects can easily occur in the orifices when the amplitude of the incident wave becomes large enough \cite{ingard1967}. 
Moreover, in many engineering applications, a grazing flow is present and its effect on the impedance of the material is inversely proportional to $\sigma$ \cite{guess1975}. For instance, it was experimentally shown \cite{auregan2016low} that the efficiency of a thin slow-sound metamaterial with $\sigma=0.023$ was divided by 100 in the presence of a grazing flow with a Mach number of 0.2. 
Therefore, in the case of high sound levels or in the presence of flow, the additional constraint of having a high open area ratio must be added in the design of structures that are thin and absorbent at low frequency. 

From this perspective, the use of elastic membranes and decorated elastic membranes as sound absorbers is an interesting option. \cite{frommhold1994,  ma2014, yang2017}. In most of the studied structures, the membrane represents a large part of the active surface and the added mass (platelets) is of smaller size. The maximum absorption appears for an hybrid resonance \cite{ma2014} due to the interaction between two modes of coupled system consisting of the membrane, the platelet and the air cavity. 
By a proper arrangement of the various parameters very impressive results can be obtained both in reflection \cite{ma2014} and in transmission \cite{wang2018acoustic}.
At the resonance frequency, the three main parameters of membrane absorbers (equivalent mass, stiffness and damping) are very sensitive to the characteristics of the membrane. By changing one of the properties of the membrane (its tension, its mass density,...) or the size of the added mass, the three equivalent parameters are simultaneously changed. 

\begin{figure}[h]
\begin{center}
\includegraphics[width=0.8\columnwidth]{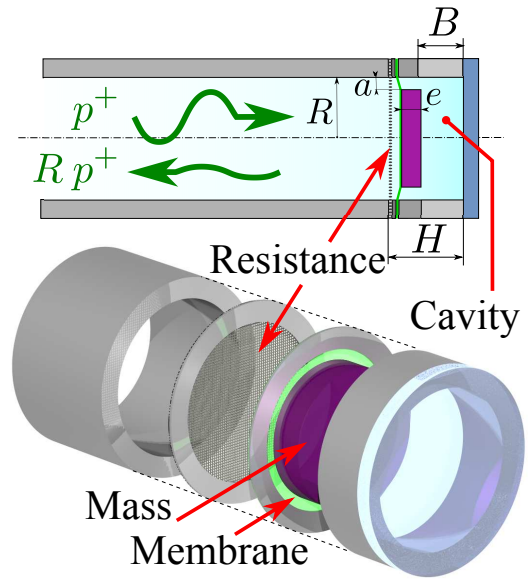}
\end{center}
\caption{\label{fig_1}Sketch of an ultra-thin low frequency (UTLF) resonator.}%
\end{figure}

This letter presents an ultra-thin low frequency (UTLF) resonator for which the three parameters (mass, rigidity and damping) can be modified independently of each other. A model has been measured and perfect absorption occurs for a sub-wavelength ratio $r_H=\lambda/H = 200$ which is, up to now, the highest sub-wavelength ratio experimentally demonstrated for a perfect absorber. 

This UTLF resonator is ideally composed of three elements displayed on Fig. \ref{fig_1}: A volume of air sealed in a cavity where the compression of the air acts as a spring, a mass that moves like a piston in the normal direction (without letting the air trapped in the cavity pass through with the help of a membrane) and a thin resistive layer that dissipates energy. 
The main difference compared to decorated elastic membranes absorbers \cite{ma2014} is that the membrane plays almost no role in the frequency of perfect absorption. The role of the very flexible membrane is just to guide the motion of the mass and seal the cavity. It will be demonstrated that the membrane also produces some damping.

As the low frequency regime is considered (any axial dimension is much smaller than $\lambda=c_0/f$ ), the continuity of the acoustic flow rate is assumed across the resistance and the mass: $v_i = v_m=v_c $ where $v_i$, $v_m$ and $v_c $ are respectively the incident wave velocity, the velocity of the mass and the mean velocity entering in the cavity. 
For simplicity, $a$, the difference between the radius of the tube and the radius of the mass, is neglected compared to the tube radius $R$. 

The UTLF resonator can be described by its impedance $Z_{\mathrm{UTLF}}$ (normalized by the air characteristic impedance $Z_0=\rho_0 c_0$ where $\rho_0$ is the air density) which is the sum of three terms since the elements are mounted in series.
The first term in the impedance is linked to the air compressibility in the cavity of thickness $B$ and can be written $Z_c =p_c/Z_0 v_c=c_0/\I \omega B= 1/\I \hat{\omega} C$ where $p_c$ is the uniform acoustic pressure in the cavity, $\omega=2 \pi f$, $\hat{\omega}=\omega/(2 \pi f_R)$ where $f_R$ is the resonance frequency of the UTLF resonator. 
The compressibility term 
$$
C = \frac{2 \pi f_R B }{c_0} =\frac{2\pi }{r_B}
$$ 
directly characterizes the sub-wavelength ratio $r_B = \lambda_R/B$ at resonance. 
This value of the cavity impedance is only valid at very low frequencies where $\omega B/c_0 \ll 1$. For higher frequencies or higher cavity thickness, a more exact value is $Z_c =1/\I \tan(\hat{\omega} C)$. 

The second term in the impedance is linked to the mass motion subjected to the pressure difference $\Delta p_m$ between its two faces. The associated impedance is $Z_m = \Delta p_m/Z_0 v_i =\I \omega m/\rho_0 c_0 S_m = \I \hat{\omega} L$ where $m=\rho_m S_m e$ is the mass of the moving part, $S_m=\pi (R-a)^2$ is the mass area, $\rho_m$ is the density of the mass material and $e$ is the mass thickness. 
It can be noted that the moving mass is connected to the fixed tube by a membrane that acts as a spring but this effect is neglected due to the low value of the membrane stiffness.
The last part of the impedance is linked to the resistive layer and is $Z_R= \Delta p_R/Z_0 v_i = R_e$ where $\Delta p_R$ is the pressure drop through the resistive layer and $R_e$ is the normalized resistance of the resistive layer. As a result, the UTLF impedance can be written $Z_{\mathrm{UTLF}}=R_e+ \I \hat{\omega} L+1/\I \hat{\omega} C$.

To be a perfect absorber when $ \hat{\omega}=1$, the UTLF impedance must perfectly match the characteristic impedance of the air. This is achieved when the two conditions $R=1$ and $LC=1$ are met. In this case, the normalized UTLF impedance can be written:
\begin{equation}
Z_{\mathrm{UTLF}}=1+\frac{1}{C}\left(\I \hat{\omega} +\frac{1}{\I \hat{\omega}} \right)
\label{eq:Z_UTLF}
\end{equation}
showing that the impedance depends only on the sub-wavelength ratio $r_B$. 
The absorption coefficient is defined by $\alpha=1-|(Z-1)/(Z+1)|^2$ and relation (\ref{eq:Z_UTLF}) shows that it is equal to 1 for $ \hat{\omega}=1$.

\begin{figure}[h]
\begin{center}
\includegraphics[width=\columnwidth]{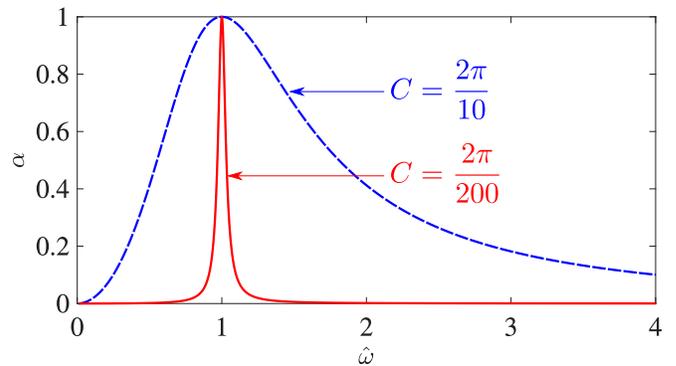}
\end{center}
\caption{\label{fig_2} \textsl{Absorption  coefficient $\alpha$ of a perfect resonant absorber as a function of the dimensionless frequency $\hat{\omega}$ for two values of the  sub-wavelength ratio: $r_B=10$ (dashed blue) and $r_B=200$ (continuous red).}}%
\end{figure}

The change rate of  $\alpha$ around $ \hat{\omega}=1$ is linked to the slope of the imaginary part of the impedance which is given by 
$|\mathrm{d}  Z_{\mathrm{UTLF}}/\mathrm{d} \hat{\omega}| = 2/C$ for $ \hat{\omega}=1$. 
Using a Taylor expansion near the resonance when $C$ is small, the frequency bandwidth $\Delta f$ for which $\alpha>0.75$ is given by 
 $\Delta f /f_R = 2 C/\sqrt{3}$. Then, $\alpha$ has an increasingly sharp peak as the sub-wavelength ratio $r_B$ increases as it is shown in Fig. \ref{fig_2}. 
In principle, there is no theoretical upper limit for the sub-wavelength ratio but the price to pay to have highly sub-wavelength cavities is to have a very small bandwidth. From a practical point of view (as will be seen later), the increase in the sub-wavelength ratio $r_B$ is limited by the losses that always become large when $r_B$ increases.

\begin{figure}[h]
\begin{center}
\includegraphics[width=\columnwidth]{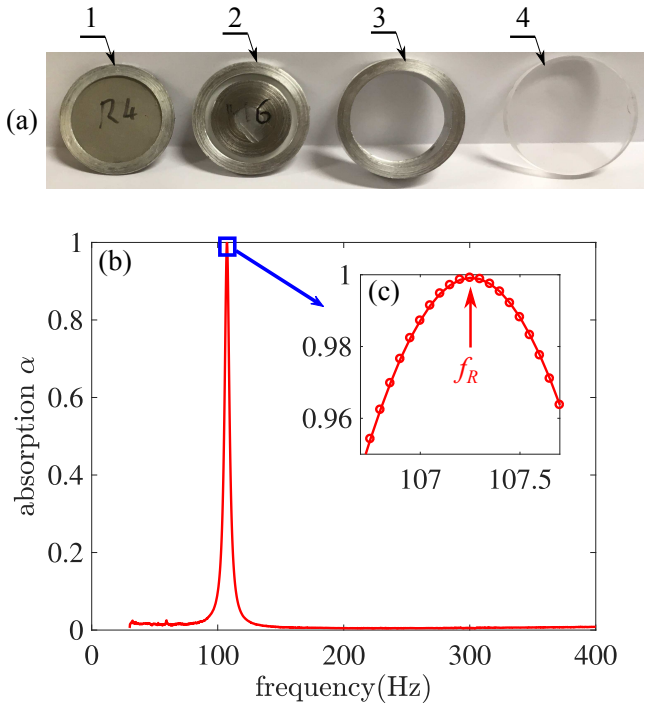}
\end{center}
\caption{\label{fig_3}  \textsl{Experimental results:(a) Picture of the elements of the UTLF: $\underline{\mathbf{1}}$ Resistive layer, $\underline{\mathbf{2}}$ Mass element, $\underline{\mathbf{3}}$ Spacer tube to make the cavity, $\underline{\mathbf{4}}$ plexiglass cover.
(b) Measured absorption  coefficient $\alpha$ of the UTLF as a function of frequency. (c) Zoom around the perfect absorption.
}}%
\end{figure}

Figure \ref{fig_3}(a) shows the different elements that composes a manufactured sample of UTLF. The first element is a resistive layer, $\underline{\mathbf{1}}$ on Fig. \ref{fig_3}(a), consisting of a metal wire mesh glued to a short tube (inner radius $R $= 15 mm, outer radius 19 mm, 1 mm long). Many wire meshes are available with different resistances and the normalized resistance of the one used here was measured at $R_e$= 0.29.  The second part is the mass element, $\underline{\mathbf{2}}$ on Fig. \ref{fig_3}(a). The mass is a steel disk ($\rho_m = 7800$ kg.m$^{-3}$)with a radius of  $R-a$ = 13 mm and a thickness of $e$ = 3 mm. A slightly tensioned latex membrane 20 $\mu$m thick is glued to a short tube (3 mm long) and then the mass is glued to this membrane. The cavity consists of tubes of various lengths $B$ and a plexiglas disk to close it, $\underline{\mathbf{3}}$ and $\underline{\mathbf{4}}$ on Fig. \ref{fig_3}(a). 

When a cavity is made with a tube of thickness $B$=12 mm, the theoretical value of the resonance frequency is given by 
$$
f_R=\frac{c_0}{2\pi}\sqrt{\frac{\rho_0 }{\rho_m e B}} = 111 \: \mathrm{Hz}.
$$
This value can be compared to the measured frequency given in \ref{fig_3}(b). A nearly perfect absorption ($\alpha=$ 0.9991) is obtained for $f_R=107.25$ Hz.
The 4\% error made on the frequency value is small considering all the simplifications that have been made in the calculations.
The value of the compressibility term is $C= 2 \pi/269$. The bandwidth for which $\alpha>0.75$ is $\Delta f$ = 2.5 Hz = $C f_R$. The total height of the material $H$ = 16 mm is the sum of the cavity height $B$, mass thickness $e$ and thickness of the resistive layer glued on its support (1 mm). Thus, the sub-wavelength ratio of this material is $r_H= \lambda/H$ = 201.6. 
Up to now, this ratio between wavelength and material thickness is the highest ever measured in an absorbent material. 

The resonance frequency is very easy to adjust since it depends only on the mass of the moving part and on the height of the cavity. 
The simplicity of the adjustment comes from the fact that membrane effects can be neglected to compute $f_R$ because of the low value of its equivalent stiffness. In the device under study, the resonance frequency of the mass element measured with the membrane alone (without the cavity) is about 35 Hz while it increases to 107.25 Hz in the presence of the cavity. This indicates that the equivalent stiffness of the membrane is 9 times lower than the one of the cavity with a thickness of $B$=12 mm.

\begin{figure}[h]
\begin{center}
\includegraphics[width=\columnwidth]{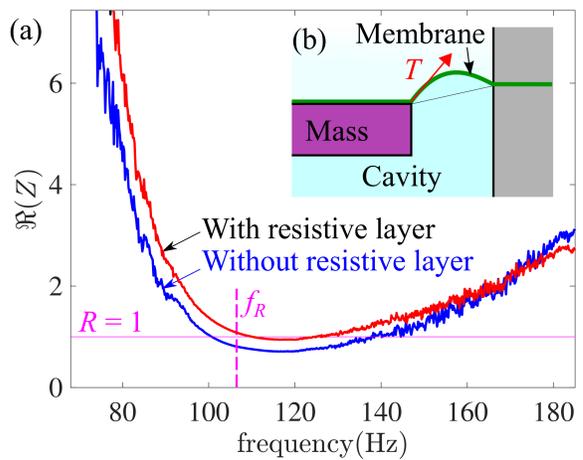}
\end{center}
\caption{\label{fig_4} \textsl{(a) Real part of the impedance (resistance) of the UTLF absorber without the resistive layer (blue) and with the resistive layer (black). (b) Schematic deformation of the membrane.
}}%
\end{figure}

The membrane is involved in the dissipative part of the impedance which is shown in Fig. \ref{fig_4}(a). 
The blue line in this figure gives the dissipative effect when the resistive layer is not present. 
In this case, the dissipative effects come solely from the membrane. 
First of all, it can be noted that this dissipation is relatively high (the minimum value of $\Re(Z)$ is 0.71 representing 71\% of the target value). It has already been noted in \cite{ma2014} that even a small dissipation in the membrane material could result in significant absorption of the overall system. 
 Second, it can be noted that the dissipation curve appears to result from two different effects. 
For $f>$140 Hz, the dissipation increases with frequency. This effect is the only one that is experimentally observed when there is no cavity. 
This dissipation can therefore be attributed to the mechanical damping resulting from the elongation of the membrane when the mass moves. 
For $f<$90 Hz a second phenomenon appears for which the resistance decreases with frequency. It has been experimentally observed that this second resistance increases as the cavity thickness decreases. One possible explanation is that, when the air trapped in the cavity becomes difficult to compress or expand (i.e. for small volume or low frequency), it inflates and aspirates  the membrane as schematized in Fig. \ref{fig_4}(b), which increases the dissipative phenomena.

The combination of the two previous effects induces a minimum of resistance for a frequency close to resonance.
When this minimum is smaller than 1, it is possible to add a resistive layer that shifts the resistance to higher values, see the red curve in Fig. \ref{fig_4}(b) in order to adapt the resistance exactly to 1 at the resonant frequency. This is what has been done to obtain the perfect absorber whose absorption curve is presented on the Fig. \ref{fig_3}(b). In the present case, the addition of a resistive layer only changes the maximum attenuation by a few percent. But for less extreme conditions ($\lambda/B \ll $200), the attenuation due to the membrane can be much lower and the addition of a resistive layer very useful to obtain a perfect absorption.

To conclude, we have seen that it is possible to build a UTLF absorber with very simple elements: mass supported by a very flexible membrane, cavity closed by the membrane and resistive layer. This device can have a large open area to minimize non-linear effects or reduce dependence on grazing flows. The frequency of the UTLF absorber can be easily adjusted by simply changing the mass or thickness of the cavity. With this type of absorber, it is possible to obtain a very high ratio between the wavelength and the thickness of the material but, as a necessary counterpart, the bandwidth is very narrow.  
The energy dissipation comes partially from the membrane and partially from the resistive layer. When trying to further reduce the size of the cavity, the resistance of the absorber becomes greater than the characteristic impedance of the air due to the dissipation in the membrane. The dissipation is therefore the limiting factor in the decrease in the size of this type of absorbers.
This remark is not specific to this absorber but can be generalized to many resonant sub-wavelength absorbers: It is the damping that limits the sub-wavelength ratio in resonant absorbing devices that are build with three in series elements (mass, spring and damping) like the classical Helmholtz resonator. 
One possibility to overcome this limit is to add some gain in the system by supplying external energy using, for example, electro-dynamical devices \cite{fleury2015invisible, lissek2018toward}, thermo-acoustic devices\cite{biwa2016experimental} or using the flow\cite{auregan2017p} which is present in many engineering applications.

\begin{acknowledgments} 
This work was supported by the "Agence Nationale de la Recherche" international project FlowMatAc No. ANR-15-CE22-0016-01.
\end{acknowledgments}

\end{document}